# RS-FME-SwinT: A Novel Feature Map Enhancement Framework Integrating Customized SwinT with Residual and Spatial CNN for Monkeypox Diagnosis

Saddam Hussain Khan[1]*, Rashid Iqbal[1]

[1]Artificial Intelligence Lab, Department of Computer Systems Engineering, University of Engineering and Applied Sciences (UEAS), Swat 19060, Pakistan

Email: saddamhkhan@ueas.edu.pk

## Abstract

Monkeypox (MPox) has emerged as a significant global concern, with cases steadily increasing daily. This serious infectious disease, characterized by skin lesions and potential complications, poses a pandemic threat in the post-COVID-19 era. Transmission occurs through direct contact, leading to diverse dermatological symptoms that demand early and accurate diagnosis. Conventional detection methods, including polymerase chain reaction (PCR) and manual examination, exhibit challenges of low sensitivity, high cost, and substantial workload. Therefore, deep learning offers an automated solution; however, the datasets include data scarcity, texture, contrast, inter-intra class variability, and similarities with other skin infectious diseases. In this regard, a novel hybrid approach is proposed that integrates the learning capacity of Residual Learning and Spatial Exploitation Convolutional Neural Network (CNN) with a customized Swin Transformer (RS-FME-SwinT) to capture multi-scale global and local correlated features for MPox diagnosis. The proposed RS-FME-SwinT technique employs a transfer learning-based feature map enhancement (FME) technique, integrating the customized SwinT for global information capture, residual blocks for texture extraction, and spatial blocks for local contrast variations. Moreover, incorporating new inverse residual blocks within the proposed SwinT effectively captures local patterns and mitigates vanishing gradients. The proposed RS-FME-SwinT has strong learning potential of diverse features that systematically reduce intra-class MPox variation and enable precise discrimination from other skin diseases. Finally, the proposed RS-FME-SwinT is a holdout cross-validated on a diverse MPox dataset and achieved outperformance on state-of-the-art CNNs and ViTs. The proposed RS-FME-SwinT demonstrates commendable results of an accuracy of 97.80%, sensitivity of 96.82%, precision of 98.06%, and an F-score of 97.44% in MPox detection. The RS-FME-SwinT could be a valuable tool for healthcare practitioners, enabling prompt and accurate MPox diagnosis and contributing significantly to mitigation efforts.

## 1. Introduction

The Zoonotic Orthopoxvirus, which causes an infectious disease called Monkeypox (MPox), belongs to the same poxviridae family as cowpox and smallpox and mainly infects monkeys and rodents, and can also spread from person to person [1], [2]. The first identification of the virus occurred in 1958, a pox-like disease affected a monkey in a Danish laboratory, and the first human infection was recorded in 1970 in the country that is known as Congo-Kinshasa, amidst initiatives to eliminate smallpox [3], [4]. The virus is mostly found in the central and western regions of Africa and can be transmitted through various ways, such as touching, biting, breathing, or contact with mucous membranes [5].

Early symptoms of MPox comprise fever, bodily ailments, and tiredness, followed by skin lesions that start as a red bump [6]. MPox is less contagious than COVID-19, but its cases have increased from 50 in 1990 to 5000 in 2020, with global outbreaks reported in 2022 [7], [8]. According to the CDC [9], 99,518 Mpox cases and 207 deaths were recorded in 2022-2023. From January 2024 to 10$^{th}$ September 2024, more than 24,000 confirmed or suspected MPox cases caused by MPox Virus clade I and clade II, resulting in over 600 deaths, have been reported across African Union Member States, with more than 5,000 of these cases confirmed, leading to heightened concern and fear on social media [10]–[14]. The lack of expert doctors and testing kits makes it difficult to manage the disease, with the current diagnostic method, polymerase chain reaction (PCR), being expensive and slow [15], [16], and no specific antiviral treatment for MPox [17]. Although there is no specific cure, the CDC has approved Brincidofovir and Tecovirimat, originally for smallpox, for MPox [18], [19]. Vaccination, FDA-approved but underused in the U.S., is practiced in other countries using smallpox vaccines against MPox [20]. Diagnosis involves manually observing skin lesions and confirming the virus through electron microscopy or PCR, commonly used in COVID-19 diagnosis [21]–[23].

Deep Learning (DL), especially deep Convolutional Neural Networks (CNNs), a valuable tool for healthcare professionals, demonstrates promise in the field of medical imaging, such as cancer, tumor, and COVID-19 detection [24]–[26] [27]. Although various DL models exist for the detection of skin diseases, their application to MPox detection remains limited [28], [29]. Studies using DL for MPox classification face challenges including model interpretability, computational costs, and limited consideration of multiclass datasets [30]–[32]. CNN facilitates the automatic extraction of features and proving useful in medical image analysis for various diseases, including COVID-19 [33]–[35]. With the increasing prevalence of MPox cases and limited testing tools, DL models emerge as a promising avenue for an automatic solution. DL presents an opportunity to mitigate challenges such as the scarcity of RT-PCR kits, potential inaccuracies in results, higher costs, and extended turnaround times commonly encountered in traditional testing methods [36]. DL models can help in developing effective triage strategies for diagnosing MPox [37]. However, diagnosing MPox from skin images presents challenges due to infected variability, texture variation, illumination, inter-intra class variation, homogeneity with other skin infectious diseases, and scarcity of labeled data [38].

Recent methods leverage advanced DL techniques, including deep CNNs and Vision Transformers (ViT), to effectively capture both global and local features, addressing key challenges in medical image analysis. These approaches employ transfer learning (TL) and optimized algorithms to mitigate data scarcity issues, particularly in emerging cases with diverse datasets, thereby enhancing model generalization and robustness. Therefore, this study introduces an innovative hybrid DL methodology as illustrated in Figure 1, termed "RS-FME-SwinT", proposed for MPox diagnosis. The proposed approach amalgamates TL-based Residual Learning and Spatial Exploitation CNN architectures with a customized Swin Transformer (SwinT) to effectively capture multi-scale correlated features. The primary contributions of the work include:

- This study introduces a novel DL hybrid "RS-FME-SwinT" approach that integrates the proposed SwinT with customized Residual and spatial CNNs to capture multi-scale global dependencies and local fine-grained correlated information for MPox disease diagnosis.
- The proposed RS-FME-SwinT framework implements feature maps enhancement (FME) of customized SwinT through TL-based residual and spatial CNN learning. Consequently, residual learning captures texture and pattern information, and spatial blocks learn minor contrast variations in local feature space. The integrated FME approach learns diverse feature maps to effectively reduce intra-class MPox variation and enables precise discrimination from skin diseases, considering distinct virulence and mortality rates.
- The inclusion of inverse residual blocks (IRB) in the customized SwinT improves local pattern extraction and mitigates vanishing gradient issues. Moreover, SwinT learns patch correlation via cross-overlapping and Multi-Head Self-Attention (MHA) captures global dependency, enhancing model performance.

- The proposed RS-FME-SwinT techniques are rigorously evaluated on a diverse MPox dataset and compared to the best state-of-the-art CNNs and ViTs. Moreover, RS-FME-SwinT demonstrates outperformance in MPox detection and effectively discriminates MPox from similar viral infections, including chickenpox, measles, and cowpox.

The subsequent sections of this article are organized as follows: Section 2 provides an in-depth review of existing research on CNN, ViT, and hybrid models for MPox detection, highlighting current gaps and limitations. Section 3 introduces the proposed MPox detection framework, while Section 4 details the dataset, experimental setup, and performance metrics employed in the study. Section 5 presents and analyzes the results, comparing the proposed framework with alternative techniques. Finally, Section 6 concludes the study and offers suggestions for future research directions.

## 2. Literature Review

MPox is a zoonotic viral disease affecting both animals and humans, closely related to the variola virus, which causes smallpox. First identified in 1970 in the Democratic Republic of Congo, MPox has since spread to multiple countries. The WHO classifies MPox as a moderate global health risk, underscoring the importance of early diagnosis for effective prevention and control. Diagnosis of MPox typically involves manual observation of skin lesions, followed by confirmation through electron microscopy or PCR, similar to methodologies used in COVID-19 diagnosis. DL has emerged as a powerful, automated tool in healthcare, showing significant potential for improving MPox diagnosis. In particular, CNNs and ViTs have proven effective in predicting MPox and other skin diseases, offering advanced capabilities in feature extraction and disease analysis.

Recent studies have explored CNNs and ViTs for MPox diagnosis, utilizing diverse algorithms, datasets, and assessment metrics. MiniGoogleNet has been utilized on the MPox Skin Lesion Dataset (MSLD) dataset, achieving 97.08% accuracy in diagnostic support for MPox detection [32]. Similarly, [39] employed CNN architectures, including ResNet18, GoogleNet, EfficientNetb0, NasnetMobile, ShuffleNet, and MobileNetV2 on the MSLD, with MobileNetV2 achieving 91.11% accuracy. Another study [40] focused on the MSLD dataset using MobileNetV2 and VGGNet, attaining 91.38% accuracy. In addition, [32] achieved 97% accuracy with MiniGoogleNet on the MSLD dataset, though the Area Under Curve (AUC) was 0.76, reflecting the binary classification nature of the task. Similarly, [41] conducted a comparative analysis of pre-trained DL models, including VGG, ResNet, InceptionV3, InceptionResNet, Xception, MobileNet, DenseNet, and EfficientNet on the MPox dataset-2022, yielding precision, recall, F1-score, and accuracy values of 85%, 85%, 85%, and 87.13%, respectively. Additionally, [31] reported good performance with MobileNetV2 in auxiliary decision support systems for hospitals, using CNNs on the MPox Skin Images Dataset (MSID). Moreover, [42] integrated deep TL methods with the convolutional block attention module, achieving an accuracy of 83.89% on MSID. While these studies primarily leverage CNNs for local feature extraction, they encounter challenges in capturing global features, leading to spatial inconsistency in smooth regions.

More recently, several studies have applied ViTs to address CNN challenges for MPox detection [43]. ViTs, originally designed for Natural Language Processing (NLP) models, have been successfully applied to various vision tasks, both lower-level and higher-level [44]. In a comparative analysis conducted by [45], DL models, including SVM, K-Nearest Neighbors (K-NN), ResNet50 with TL, and ViT, were evaluated on the MSLD dataset. ViT demonstrated notable performance with an accuracy of 93%, precision of 93%, recall of 91%, and an F-score of 92%, showcasing optimal performance over SVM and K-NN in MPox disease diagnosis. Furthermore, [46] subjected ViT to the MSLD dataset, achieving precision, recall, and F1 score values of 95%, along with an accuracy of 94.69%, highlighting the robustness of ViT in different diagnostic settings.

- CNNs excel at extracting local features, potentially reducing spatial correlation and impacting performance on larger, more complex patterns.
- ViTs segment images into linear patches, making them sensitive to patch size. This approach may inadequately capture local and low-resolution details and cross-window connections.
- The vanishing gradient problem has historically challenged the training of DL models, as gradients often diminish during backpropagation through multiple layers, leading to ineffective learning.

## 3. Methodology

The proposed methodology encompasses robust data preprocessing techniques and introduces the RS-FME-SwinT technique, designed to boost the customized SwinT channels through residual learning and spatial CNNs. Our proposed approach excels in learning discriminative features by amalgamating aforementioned components that effectively mitigate intra-class variation inherent in MPox cases while facilitating precise differentiation from other skin diseases (inter-class). Finally, comprehensive comparisons are conducted against state-of-the-art ViTs and CNNs to validate the efficacy of our proposed techniques. The overall framework for MPox detection is elucidated in the accompanying Figure 1, providing a visual representation of our approach's technique and workflow.

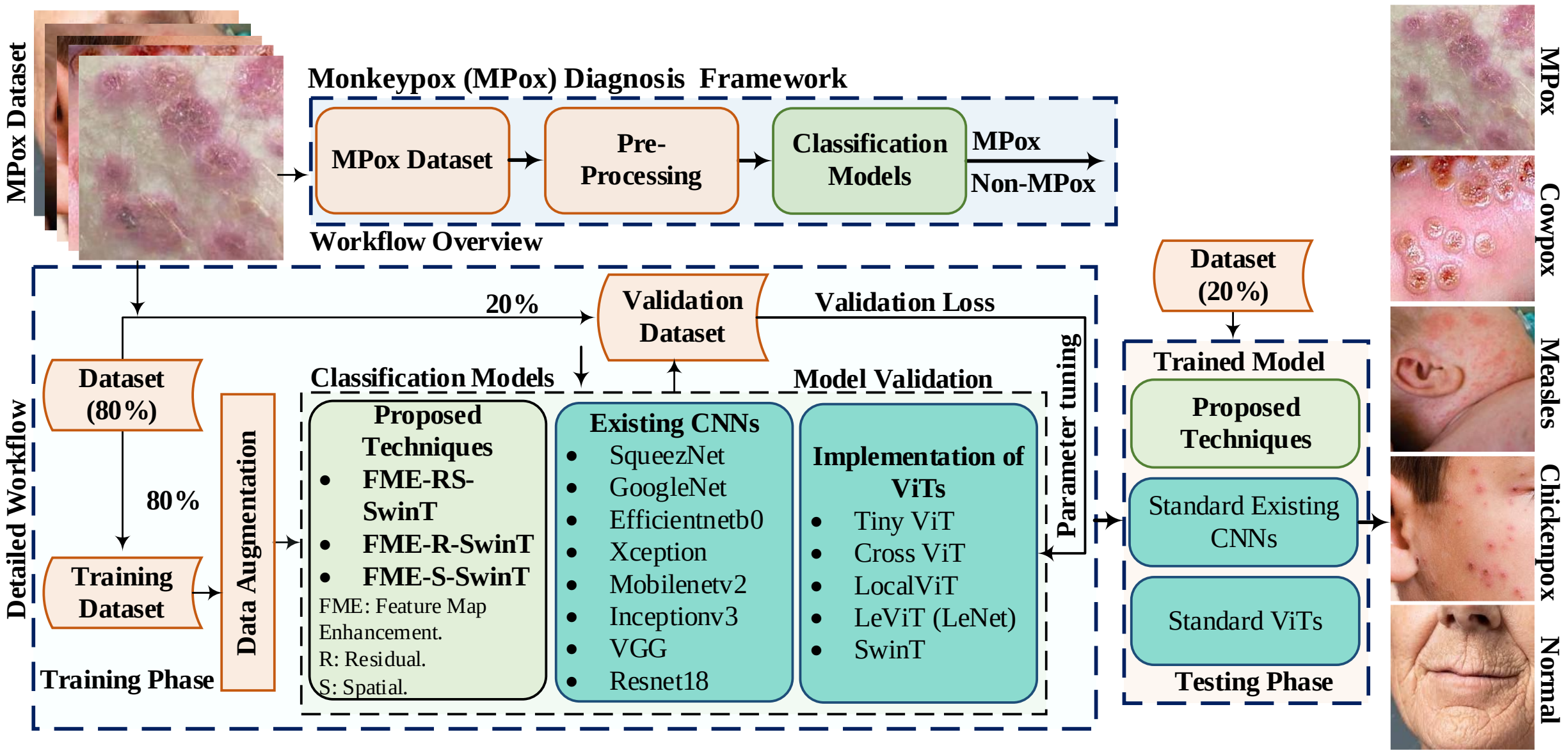

Figure 1. Monkeypox Diagnosis framework.

### 3.1.Pre-processing

Efficient data preprocessing aids in mitigating overfitting by facilitating improved model learning and generalization capabilities. The data preprocessing encompassed fundamental operations including resizing and data augmentation. The dataset underwent transforming image dimensions to 224 × 224 pixels. Moreover, data augmentation (DA) is a standard technique used to address overfitting, a common challenge in DL models, particularly on limited datasets as shown in Table 1. DA is a technique that produces more data by applying slight changes to synthetic data derived from the original. DA technique is useful in the initial stages of MPox data collection to get an increased, diverse, and stringent high-quality dataset. In this regard, the Keras Imageaugmentor library is used that allow various image DA like rotation, shifting, and flipping. Algorithm 1 presents the pseudocode of the DA methods employed.

**Algorithm 1** Data Augmentation Pseudo-Code

Input: load original images x using OpenCV.

Resize each image to 224 × 224 pixels.

Store resized images as arrays in a list.

Invoke Image data generator function.

For $n$ from 1 to 20 do.

Set batch size to 16.

Save images to a folder.

Save images as JPG files.

End for.

**End of Algorithm**

Table 1. Data Augmentation detail.

| **Parameter** | **Values** |
|---|---|
| Rotation | [± 30] degree |
| Shearing | [0 30] |
| Scaling | [1 1.5] |
| Translation | [±5] |
| X-Y Reflection | [±1] |

### 3.2.The Proposed RS-FME-SwinT

This study presents a novel approach to detecting MPox disease by integrating a proposed SwinT with specialized CNNs. The proposed RS-FME-SwinT leverages a novel feature map enhancement (FME) technique at the target level, comprising three key components: a customized SwinT technique and residual and spatial CNN blocks. The proposed SwinT demonstrates enhanced efficiency in capturing comprehensive global information, while the residual blocks are meticulously optimized for extracting intricate texture and pattern features. Additionally, the spatial blocks are proficient in fine-tuning local contrast. Moreover, the customized SwinT introduces an innovative residual block designed to mitigate the vanishing gradient problem through skip connections, thereby ensuring robust gradient flow. Concurrently, the spatial blocks leverage systematic 3x3 convolutional filters and implement both max and average pooling to focus on the extraction of homogeneous features and precise boundary delineation of local features. This strategic combination of global and local feature extraction enhances overall performance, efficiency, and applicability across diverse strengthened challenging MPox datasets. The key components of the proposed RS-FME-SwinT in Figure 2 are mentioned below.

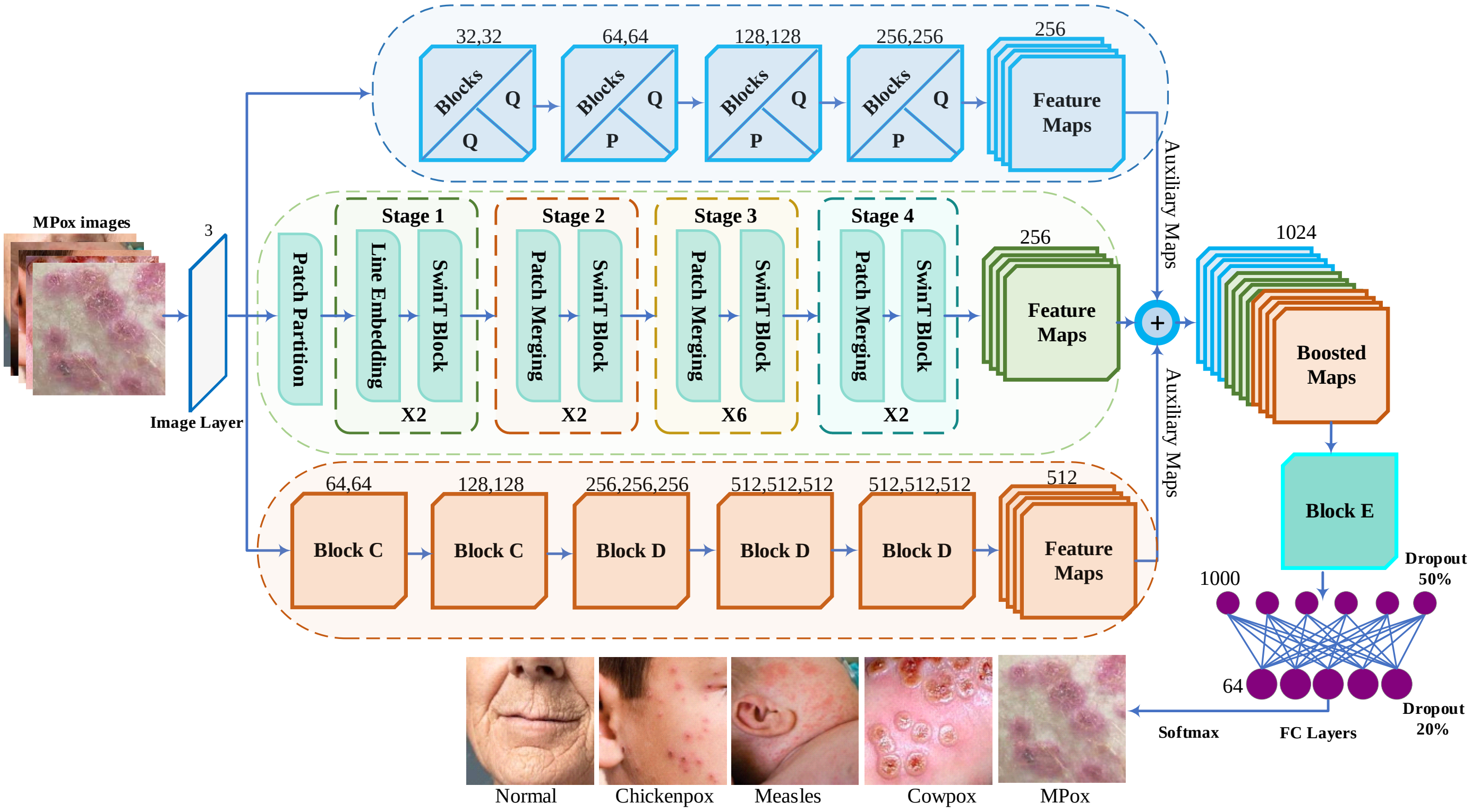

Figure 2. An overall framework for MPox detection.

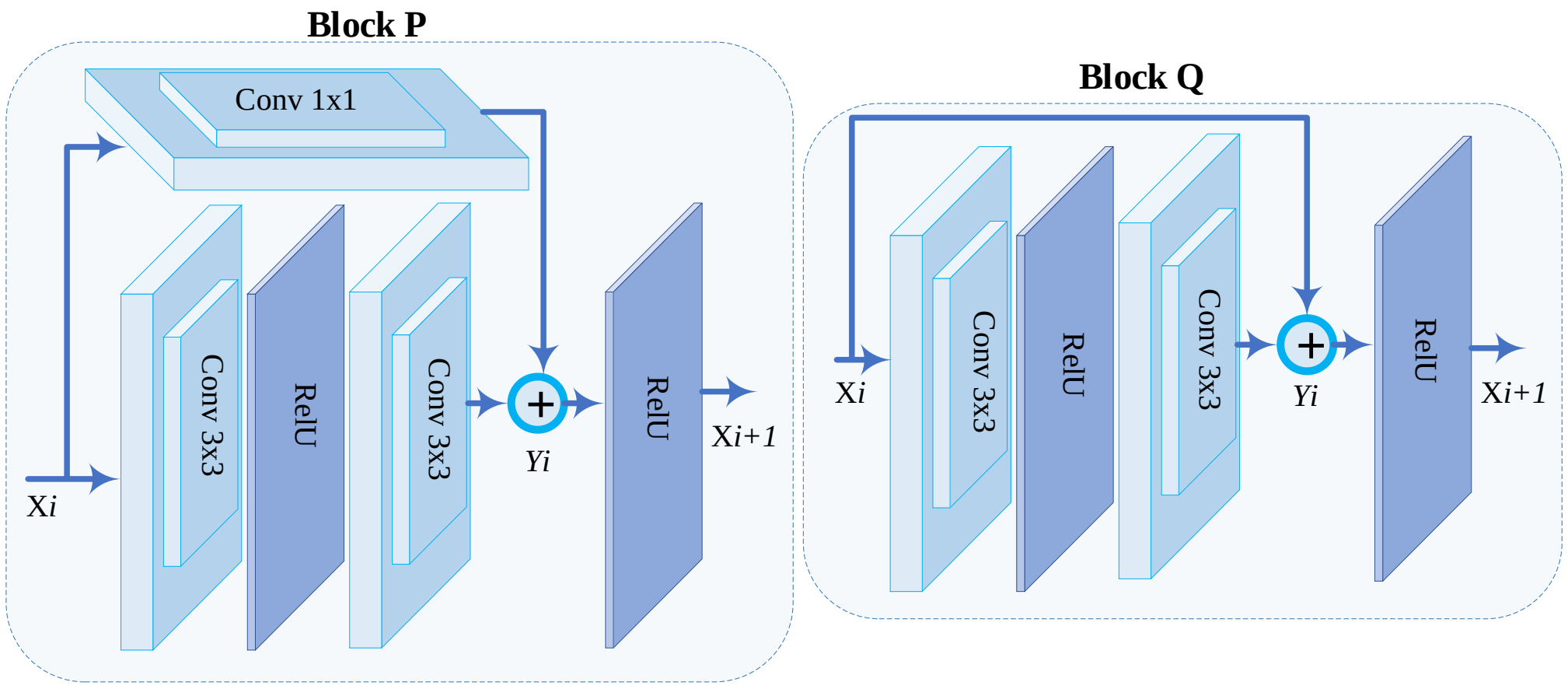

Figure 3. Overview of Residual Learning Blocks.

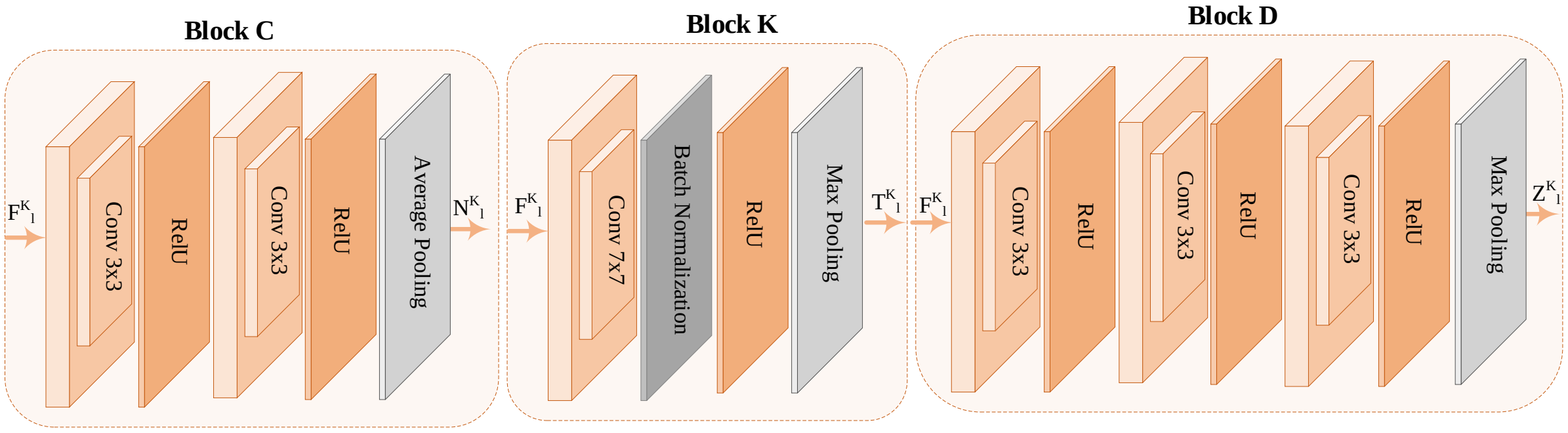

Figure 4. Spatial Exploitation Block.

#### 3.2.1. Proposed Swin Transformer (SwinT)

The proposed SwinT is specifically tailored to overcome challenges in ViT architecture from language to vision, providing distinctive advantages in a range of medical diagnosis tasks [47]. The transformer-based approach uses a hierarchical structure and divides the input images into non-overlapping patches for further processing. Consequently, the incorporation of shifted windows represents an advancement in ViT that addresses the non-overlapping local windows constraint of self-attention while maintaining connections across windows. The proposed SwinT model consists of three essential components: Patch and Positional Embedding, MHA to capture that captures global dependencies, and introduced a new IRB block. The IRB block in the customized SwinT integrates concepts from new convolutional skip connections for capturing local correlated features and reducing the chance of vanishing gradient. The SwinT embeds input images that excel in learning spatial correlations, employs self-attention, and IRB blocks to linearly project captured long-range dependencies and locally correlated features of complex lesion patterns, respectively, as illustrated in Figure 5.

#### i. Patch and Positional Embedding

The customized swinT adapts 2D images and segments into a series of flattened 2D patches, with each image patch representing a token. The encoder learns the patch relationship and produces a final feature vector by adding a special token at the sequence's beginning. These tokens serve as the image representation and undergo self-attention within the encoder, resulting in an output that is then directly routed to the classification head for categorization.

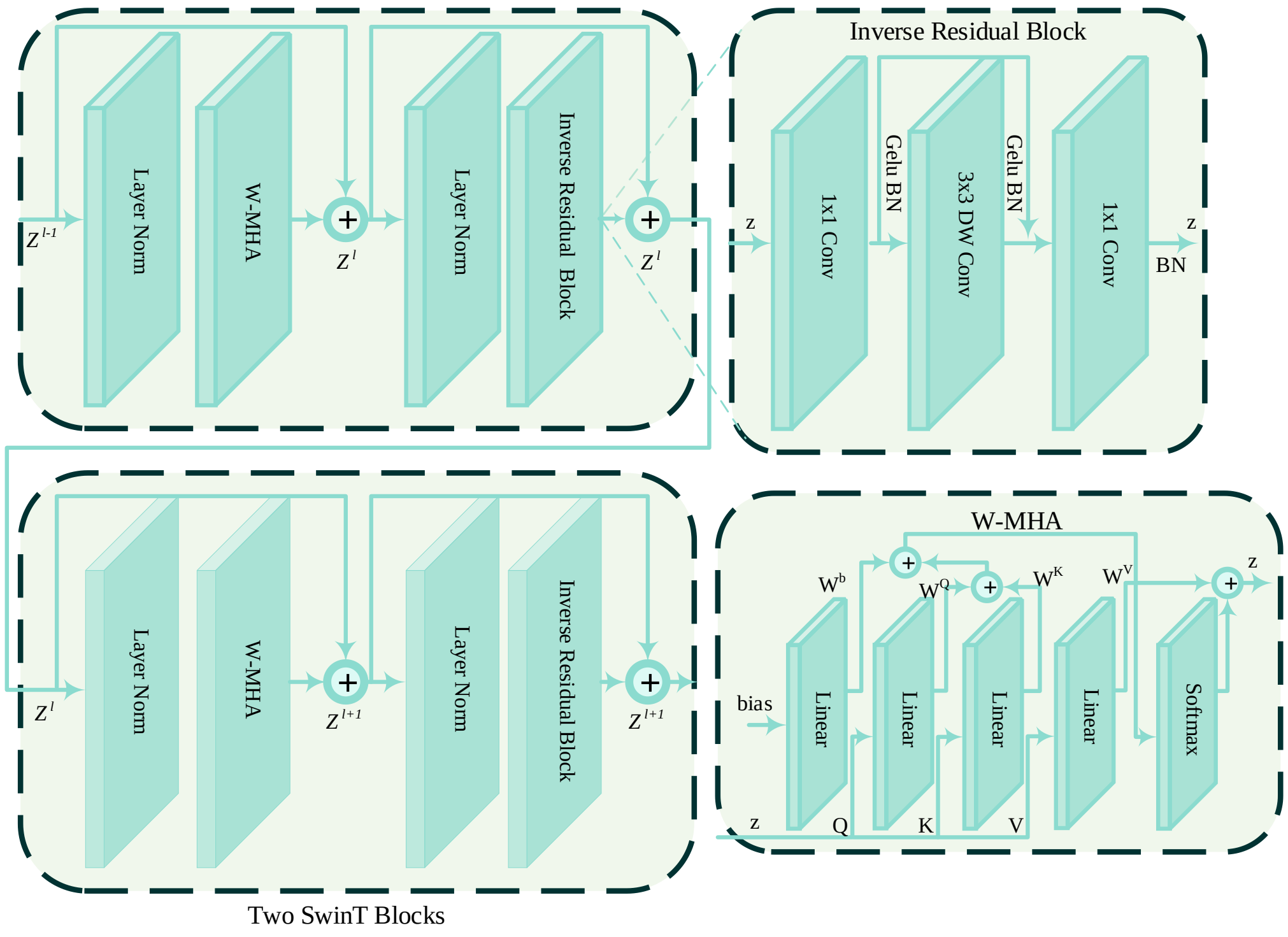

Figure 5. Customized Swin Transformer (SwinT) Block.

The dimensions of the original image are represented as A * B * C, which undergoes reshaping into $N^{th}$ patches, with $N =.\frac{A}{P} * \frac{B}{P}$. Here, (H, W) represents the height and width of the authentic image, C denotes the channel counts, and P denotes the resolution of each image patch. The SwinT resizes MPox lesion images to 224 × 224 × 3, segmenting them into 16 × 16 patches, flattened and linearly projected into high-dimensional 196 × 16 feature vectors through a transformer encoder.

$$\boldsymbol{X}_{Patch}^{N*D} = R(\boldsymbol{I}_{image}^{A*B*C}) \quad (1)$$

$$Z_0 = \boldsymbol{X}_{class}\,; \boldsymbol{X}_p^1 E; \boldsymbol{X}_p^2 E; \ldots; \boldsymbol{X}_P^N E] + E_{pos,} \quad (2)$$

$$E \in R^{(p2*c)xD}, E_{pos} \in R^{(N+1)xD} \quad (3)$$

The reshaping procedure is detailed in Eq. (1 to 3), with the parameters being specified as $N =.\frac{A}{P} * \frac{B}{P}$ and $D = p^2 * C$ depicts patch dimensions, with C denoting the count of channels. A trainable embedding is introduced into the sequence of embedded patches, akin to the (class) token in (Bidirectional Encoder Representations from Transformers) BERT. The classification head is connected to ($Z_o$) during fine-tuning swinT.

### ii. Multi-Head Self-Attention

The MHA addresses Self-Attention's tendency to focus on limited optimal positions important in ViT for explicit sequence entity representation. Various ViT architectures adapt SA but lack global attention and dependencies. In this regard, MHA is employed that integrate dense global attention mechanisms and sparsity for global-level dependencies by stacking SA blocks and therefore, enhancing effectiveness. The MHA mechanism is the central part of the Transformer and fuses input tokens with weighted fusion. MHA handles self-attention constraints by calculating multiple similarity scores in distinct subspaces and reducing computational costs while increasing expressiveness and interpretability. Moreover, MHA resembles the subspace mechanism in convolution and enables the model to extract multi-subspace features effectively.

The transformer block comprises normalization, a residual connection, an MHA with key (K), query (Q), and value (V) computations, and an IRB with skip connections. The MHA process transforms the input vector into a query (Q), key (K), and (V) vectors, applying self-attention individually to each patch. The SoftMax activation is after the dot product of Q and K, producing an attention matrix. The resulting attention scores are multiplied by V, generating the self-attention result (SA (Q, K, V)), as depicted in Eq. (4), where $d^k$ denotes the dimension of vector K.

$$SA(Q,K,V) = SoftMax\left(\frac{QK^T}{\sqrt{d^k}}\right) V \tag{4}$$

The output of the MHA block ($\mathrm{MHA}_{\mathrm{out}}$) is calculated using Eq. (5), where $\mathbf{X}_{\mathrm{in}}$ denotes the input of the transformer blocks, NORM represents the normalization layers, and MHA stands for multi-head self-attention. Subsequently, the output of the transformer blocks ($\mathrm{TF}_{\mathrm{out}}$) is computed with Eq. (6), incorporating the output of MHA.

$$MHA_{\mathrm{out}} = MHA(NORM(\boldsymbol{X}_{\mathrm{in}})) + \boldsymbol{X}_{\mathrm{in}} \tag{5}$$

$$TF_{\mathrm{out}} = IRFFN(NORM(MHA_{\mathrm{out}})) + MHA_{\mathrm{out}} \tag{6}$$

### iii. New Inverse Residual Block (IRB)

A new IRB is introduced in customized swinT that integrates residual connections and inverted bottleneck. IRB enhances the capability of feature extraction and representation in the customized swinT. Moreover, the new residual block, featuring a skip connection integrating the layer input to its output, addresses vanishing gradients and ensures seamless information flow. Furthermore, the inverted bottleneck block strategically boosts channels, applies depth-wise convolution, and then reverts to original counts, thus reducing computational costs while enhancing feature diversity. This configuration enables the IRB to effectively capture information at both local and global scales within intermediate features, elevating the overall capacity for representation. The customized IRB, operating as a Feedforward Network (FFN), consists of two inverted bottleneck blocks featuring residual connections. The initial FFN introduced in ViT [44] comprises two linear layers with an intervening GELU activation [48], replaced by a new IRB.

The proposed IRB resembles inverted residual blocks, involving depth-wise convolution, expansion, and projection layers to facilitate localized information extraction with minimal computational cost. A shortcut connection has been added to enhance the performance with batch normalization included after GELU activation and in the final linear layer, following a similar approach.

$$FFN(\boldsymbol{X}) = GELU(\boldsymbol{X}\boldsymbol{W}_1 + b_1)\boldsymbol{W}_2 + b_2 \tag{7}$$

$$\mathrm{IRB}(\mathbf{X}) = Conv(f(Conv(\mathbf{X}))) \tag{8}$$

$$\mathrm{f}(\mathbf{X}) = DWConv(\mathbf{X}) + \mathbf{X} \tag{9}$$

The first layer increases the dimension by fourfold, followed by the subsequent layer decreasing the dimension proportionally, as depicted in Eq. (7-9). Here, $\mathrm{W}_1 \in \mathrm{R}^{\mathrm{d} * 4\mathrm{d}}$ and $\mathrm{W}_2 \in \mathrm{R}^{4\mathrm{d} * \mathrm{d}}$ represent the weights of the two linear layers, and $\mathrm{b}_1$ and $\mathrm{b}_2$ are the biased terms.

### 3.2.2. Residual Learning CNN

Residual learning mitigates performance degradation in DL models through shortcut connections, enabling direct mapping and improving learning capacity [49]. Residual blocks address the vanishing gradient problem by preventing the model from getting stuck in suboptimal minima and enhancing convergence [50]. Furthermore, residual blocks actively contribute to capturing essential disease lesion textures and intricate patterns, ultimately improving detection performance. We have utilized a stacking technique, merging residual learning with four P and Q blocks to systematically enhance learning (see Figure 3). The convolutional kernel size of 3 × 3 in both P and Q blocks achieves a local receptive field (Eq. 10). In a convolutional layer, individual neurons are represented by a collection of convolutional kernels, each segmenting the input image into compact receptive fields. Here, $i_c(x, y)$ represents an element within the input image tensor $l_c$, and $e_l^k$ denotes the position of the $k^{th}$ convolutional kernel in the $l^{th}$ layer. The resultant feature map generated by the $k^{th}$ convolutional operation is denoted as $F_l^k = [f_l^k(1,1), \ldots, f_l^k(p, q), \ldots, (f_l^k(P, Q)]$. To mitigate concerns associated with internal variations in feature maps, Batch Normalization is utilized. The conversion of a feature-map $(F_l^k)$ to a normalized feature-map $(N_l^k)$, through batch normalization is formulated in Eq. (11). Here, $\mu_B$ and $\sigma_B^2$ represent the mean and standard deviation of a feature map within a small batch, respectively. The parameter $\epsilon$ is a small constant added for numerical stability. Additionally, the activation function is employed to handle non-linearity in Eq. (12). In this equation, $(F_l^k)$, denotes the output after passing through the activation function $g_a$ (.). Furthermore, down-sampling or pooling consolidates information within receptive zones, extracting the most significant response (Eq. (13)). Here, $Z_l^k$ represents the consolidated feature-map of the $l^{th}$ layer corresponding to the $k^{th}$ input feature map $(F_l^k)$, with $g_p$ (.) defines the pooling operation. Stacking TL-based residual blocks strategically explore the effective diverse feature spaces.

$$f_l^k(p, q) = \sum_c \sum_{x,y} i_c(\mathbf{X}, \mathbf{Y}) \,.\, e_l^k(u, v) \tag{10}$$

$$N_l^k = \frac{(F_l^k) - \mu_B}{\sqrt{\sigma_l^k + \varepsilon}} \tag{11}$$

$$T_l^k = g_a(F_l^k) \tag{12}$$

$$Z_l^k = g_p(F_l^k) \tag{13}$$

The residual block which applies to every few stacked layers is expressed as $y(\mathbf{X}) = F(\mathbf{X}, [\mathbf{W}_i]) + \mathbf{X}$, which denotes the residual mapping targeted for learning. Where $(\mathbf{X})$ is the input, $(F(\mathbf{X}))$ is the output of the convolutional or pooling layers, and $(y(\mathbf{X}))$ is the final output (Eq. 14). For a two-layer block, F is $\mathbf{W}_2\sigma(\mathbf{W}_1\mathbf{X})$, where $\sigma$ is activation [29], and biases are ignored.

$$y = F(\boldsymbol{X}, [\boldsymbol{W}_i]) + \boldsymbol{X} \tag{14}$$

$$y = F(\boldsymbol{X}, [\boldsymbol{W}_i]) + \boldsymbol{W}_s\boldsymbol{X} \quad (15)$$

Pixel-wise convolution, with a 1 × 1 kernel size in the P block, facilitates inter-feature map communication and transforms convolution feature maps into distinct output dimensions (Eq. 15). The block performs linear transformation with a shortcut connection denoted by $\mathbf{W}_s$ and an element-wise addition, followed by σ(y). This systematic arrangement, comprising four sequential residual blocks, helps acquire a wide range of essential features. Channels are gradually increased from 64 to 256, fortifying the learning process and ensuring meticulous refinement for superior outcomes.

### 3.2.3. Spatial Exploitation CNN

Spatial blocks are employed to systematically learn the inter-class contrast and local homogeneous variation of MPox. These blocks facilitate hierarchical feature extraction by employing multiple convolutional layers and capturing homogeneity. Each Spatial block employs a small 3x3 convolutional operation, followed by batch normalization, activation, and multiple layers for efficient local feature extraction (Eqs. 11-13) and in Figure 4. This strategic implementation enhances the network's capability to identify diverse features within skin images, thereby improving detection performance and accuracy. Each spatial block contains 3x3 convolutional layers with a padding of 1, followed by a 2x2 max-pooling layer with a stride of 2. These blocks handle spatial data across multiple layers, using 3x3 convolutional filters and both max and average pooling to improve the extraction of consistent features and the precise identification of local feature boundaries [51]. The convolutional kernels preserve spatial homogenous resolution while pooling layers capture high-intensity contrast information. Moreover, the pooling operation reduces spatial resolution at each block and increases robustness [52]. Five spatial blocks are stacked and employ repeated sequences of convolutional and pooling layers, forming a hierarchical feature extraction. This facilitates learning complex features at higher layers while preventing information loss at lower layers. Spatial blocks increase network depth for learning local homogeneous and correlated features, reduce parameters and computational complexity with small kernels, and enhance generalization through regularization.

### 3.2.4. Feature Map Enchancment Design

The proposed RS-FME-SwinT extracts diverse information from images, including global, texture, and local image representations by concatenating customized SwinT, residual, and spatial feature maps (Eq. 16) [24]. The customized SwinT employs a window-based self-attention mechanism, allowing it to detect subtle contrast variations in intensity and novel IRB block for the texture, which might be crucial for identifying MPox. Additionally, the SwinT channels are integrated with spatial and residual auxiliary feature maps generated through TL, thereby broadening the learning capacity. A deliberate boosting strategy, characterized by a gradual increment in the number of channels, ensures comprehensive and refined learning. The individual learners make decisions by analyzing diverse, image-specific patterns within the boosting process, thereby enhancing the quality of outcomes.

$$\boldsymbol{X}_{Boosted=}b(\,\boldsymbol{X}_{R}||\,\boldsymbol{X}_{S}||\,\boldsymbol{X}_{SwinT}\,) \tag{16}$$

$$\boldsymbol{X}_{DBF} = \sum_{a}^{A}\sum_{b}^{B} v_a\, \boldsymbol{X}_{Boosted} \tag{17}$$

$$\sigma(\mathbf{X}) = \frac{e^{x_i}}{\sum_{i=1}^{c} e^{x_c}} \tag{18}$$

In Eq. (16), the RS-FME-SwinT denoted as R, S, and SwinT utilize feature maps represented by $\mathbf{X}_{R}$, $\mathbf{X}_{S}$, and $\mathbf{X}_{SwinT}$ respectively. Moreover, in certain instances, SwinT channels are combined with additional generated Residual and Spatial learning channels using TL. The boosting process, represented as b(.), plays a vital role in learning diverse feature maps. Furthermore, the output undergoes refinement through a global average pooling layer, which reduces dimensionality while retaining spatial information. The final classification is conducted through a fully connected layer with softmax activation, tailored to dataset classes, including MPox, chickenpox, measles, cowpox, and normal skin. The SwinT encompasses densely fully connected layers with dropout regularization to capture and preserve target-specific characteristics while mitigating overfitting [53], [54]. Eq. (17) features the count of neural units denoted as v, and Eq. (18) includes the softmax function represented as c, associated with the class count.

### 3.2.5. The Significance of TL

TL leveraging knowledge from existing methods to tackle MPox detection. Various TL strategies, such as instance-based, parameter exploitation-based, channel-oriented, and relation-knowledge-based TL, as discussed by [55], are employed. Channel-oriented TL, in particular, offers advantages in MPox classification and pattern detection. It involves adapting pre-existing CNNs to the specific target domain through fine-tuning techniques, commonly referred to as domain adaptation [56]. This approach entails adapting features learned from one domain to MPox imaging detection tasks through fine-tuning, bypassing the requirement for extensive calibration and hyper-parameter selection as presented in Table 2. Consequently, TL offers substantial time savings during training [57].

Table 2. Hold-out based optimal hyper-parameter selection.

| Hyper-parameter | Values |
|---|---|
| Learning-rate (α) | $10^{-3}$ |
| Optimizer | SGD |
| Epoch | 10 |
| Momentum | 0.90 |
| Loss | Cross Entropy |

## 4. Experimental Setup

### 4.1.Dataset

Prompt diagnosis of MPox is imperative due to the rapid spread of the disease across countries. DL systems have the potential to streamline the diagnosis process for healthcare professionals, but their effectiveness relies on the availability of reliable data. Notably, constructing a dataset for a novel pandemic poses significant challenges, as highlighted by [58], [59]. To address this challenge, we undertook the collection of MPox patient data from diverse sources, including platforms like Kaggle [60], iStock, newspapers, and publicly available samples obtained through Google searches.

Table 3. Benchmarked Monkeypox Detail.

| Characteristics | Overview |
|---|---|
| Total | 8354 Samples |
| Measles | 770 Samples |
| Chickenpox | 1050 Samples |
| Cowpox | 924 Samples |
| Monkeypox | 4014 Samples |
| Normal | 1596 Samples |
| Train, Validation (80%) | (5348, 1337) |
| Test (20%) | (1669) |
| Image Size | 224 x 224 x 3 |

We employed holdout cross-validation in response to the constraints posed by a limited dataset. The dataset was divided, with 80% allocated for training and the remaining 20% for testing. This consolidated dataset encompasses five distinct classes: MPox, chickenpox, measles, cowpox, and normal as illustrated in Figure 6. Notably, this dataset stands as a pioneer, incorporating a substantial number of samples for each class. Specifically, we collected 1050 image samples for the chickenpox class, 770 for measles, 4014 for MPox, 924 for cowpox, and 1596 for the normal class as shown in Table 3. The primary objective of this dataset is to differentiate MPox cases from other related non-MPox cases.

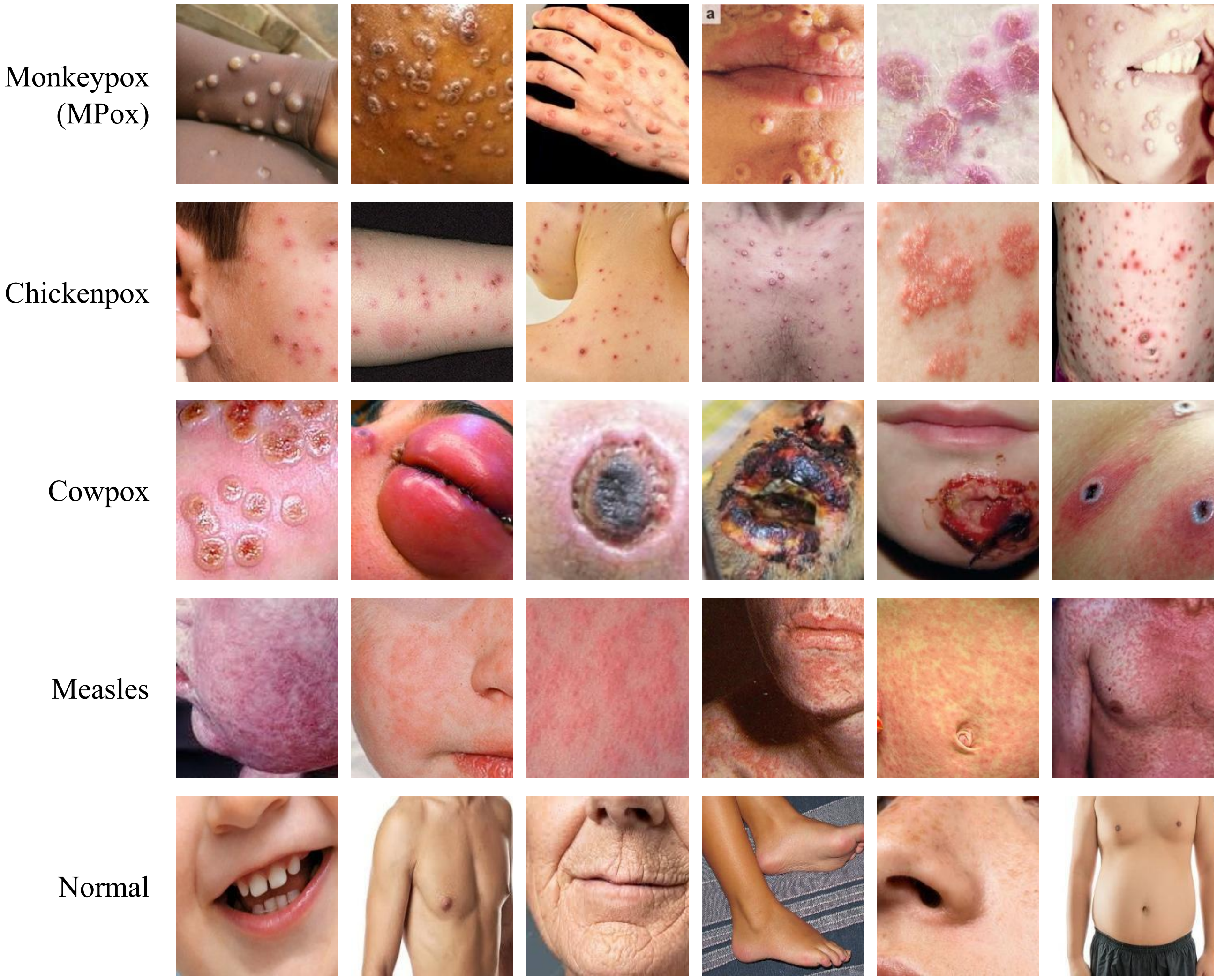

Figure 6. Dataset labeled images data comprises five classes: MPox, Chickenpox, Cowpox, Measles, and Normal.

## 4.2.Performance Assessment Metrics

Various performance metrics are employed to assess the effectiveness of the technique, depending on the nature of the data. In the case of imbalanced datasets, precision, sensitivity, and the AUC-PR curve are employed. Additionally, the F1-score is used to provide an unbiased estimate, considering both True positive (TP), True negative (TN), False negative (FN), and False positive (FP) across MPox-positive and negative classes. The negative class includes Measles, Chickenpox, Cowpox, and normal individuals. Qualitative measures encompass accuracy, sensitivity, precision, and F score, Eq's (19)–(24) show the mathematical expressions. Accuracy is the proportion of correctly classified samples in both MPox-positive and negative categories. Sensitivity and specificity are the ratios of correctly identified MPox-positive and negative patients. Precision is the ratio of accurate predictions to the total MPox positive predictions. Eq. (24) delineates the Standard Error (S.E) for sensitivity within a 95% Confidence Interval (CI). The objective is to enhance the TP rate and reduce the incidence of FN's in the detection of MPox [44], The value of z is set at 1.96 to represent the S.E. within a 95% CI.

$$\text{ACC} = \frac{\text{True positive of MPox } (TP) + True\ Negative\ of\ MPox(TN)}{(TP + TN + FP + FN)} \times 100 \tag{19}$$

$$\text{Sen} = \frac{\text{True positive of MPox } (TP)}{\text{Total positive Images of MPox } (TP + FN)} \times 100 \tag{20}$$

$$\text{Spec} = \frac{\text{True negative of MPox } (TN)}{\text{Total negative of MPox Images } (TN + FP)} \times 100 \tag{21}$$

$$\text{Pre} = \frac{\text{True positives of MPox } (TP)}{\text{True positives of MPox } (TP) + \text{ False positives of MPox } (FP)} \times 100 \tag{22}$$

$$F - \text{score} = 2\, x \frac{\text{Pre} \times \text{Sen}}{\text{Pre} + \text{Sen}} \tag{23}$$

$$C.I = z \sqrt{\frac{error(1 - \text{error})}{\text{Total Samples}}} \tag{24}$$

### 4.3.Implementation Setup

The optimization criterion for training deep CNN and SwinT models involved Stochastic Gradient Descent (SGD) using cross-entropy loss. Class probabilities were derived through the utilization of softmax, complemented by the implementation of a learning rate scheduler based on Piecewise adjustment. The learning rate was fixed at 0.0001, accompanied by a momentum of 0.95. Batch sizes of 16, and the models underwent training for 50 epochs. Subsequently, the models were trained on multiple datasets and evaluated on their respective unobserved test sets. Additionally, the deep CNN and SwinT models have been constructed using MATLAB 2023b and its DL library.

The experiments were executed on an NVIDIA GeForce GTX Titan X computer with 64 GB of RAM, leveraging CUDA technology. The average training time per epoch on the NVIDIA GeForce GTX Titan X ranged from 1 to 2 hours.

## 5. Results and Discussion

The proposed RS-FME-SwinT introduces for MPox diagnosis technique featuring customized SwinT, Residual learning, and Spatial exploitation. This technique enhances the model's efficiency in capturing global, texture, and local features from the MPox dataset, emphasizing its effectiveness in learning intra and interclass variation homogeneity, and textural variation, leading to enhanced MPox detection. The technique demonstrates superior performance compared to other state-of-the-art CNNs and ViTs in MPox detection, achieving accuracy, sensitivity, precision, and F-scores of 97.80%, 96.82%, 98.06%, and 97.44%, respectively (Table 4) and shown in the Figure 7 confusion matrix.

Table 4. Performance Analysis of the proposed techniques and existing CNNs/ViTs.

| **Model** | **Accuracy** | **Sensitivity** | **Precision** | **F-Score** |
|---|---|---|---|---|
| **Existing CNNs** | | | | |
| SqueezNet | 85.81 | 82.42 | 86.78 | 84.55 |
| GoogleNet | 86.17 | 88.02 | 82.14 | 84.98 |
| Efficientnetb0 | 91.46 | 90.95 | 90.18 | 90.56 |
| Xception | 93.21 | 92.90 | 91.61 | 92.25 |
| Mobilenetv2 | 94.41 | 94.07 | 93.17 | 93.62 |
| Inceptionv3 | 94.77 | 94.06 | 94.28 | 94.17 |
| **Existing ViTs** | | | | |
| Tiny ViT | 90.08 | 89.90 | 87.45 | 88.66 |
| Cross ViT | 92.18 | 91.17 | 91.67 | 91.42 |
| LocalViT | 92.96 | 91.37 | 92.42 | 91.89 |
| LeViT (LeNet) | 94.83 | 94.52 | 93.47 | 93.99 |
| **Proposed Setup** | | | | |
| Customized VGG | 95.13 | 94.19 | 94.34 | 94.27 |
| Customized resnet18 | 95.31 | 94.98 | 94.92 | 94.95 |
| SwinT | 95.43 | 94.99 | 94.93 | 94.96 |
| **Proposed SwinT** | 95.71 | 95.35 | 95.32 | 95.34 |
| **Proposed SwinT + Spatial** | 96.09 | 95.36 | 95.92 | 95.64 |
| **Proposed SwinT + Residual** | 97.17 | 96.57 | 97.34 | 96.95 |
| **Proposed RS-FME-SwinT** | 97.80 | 96.82 | 98.06 | 97.44 |

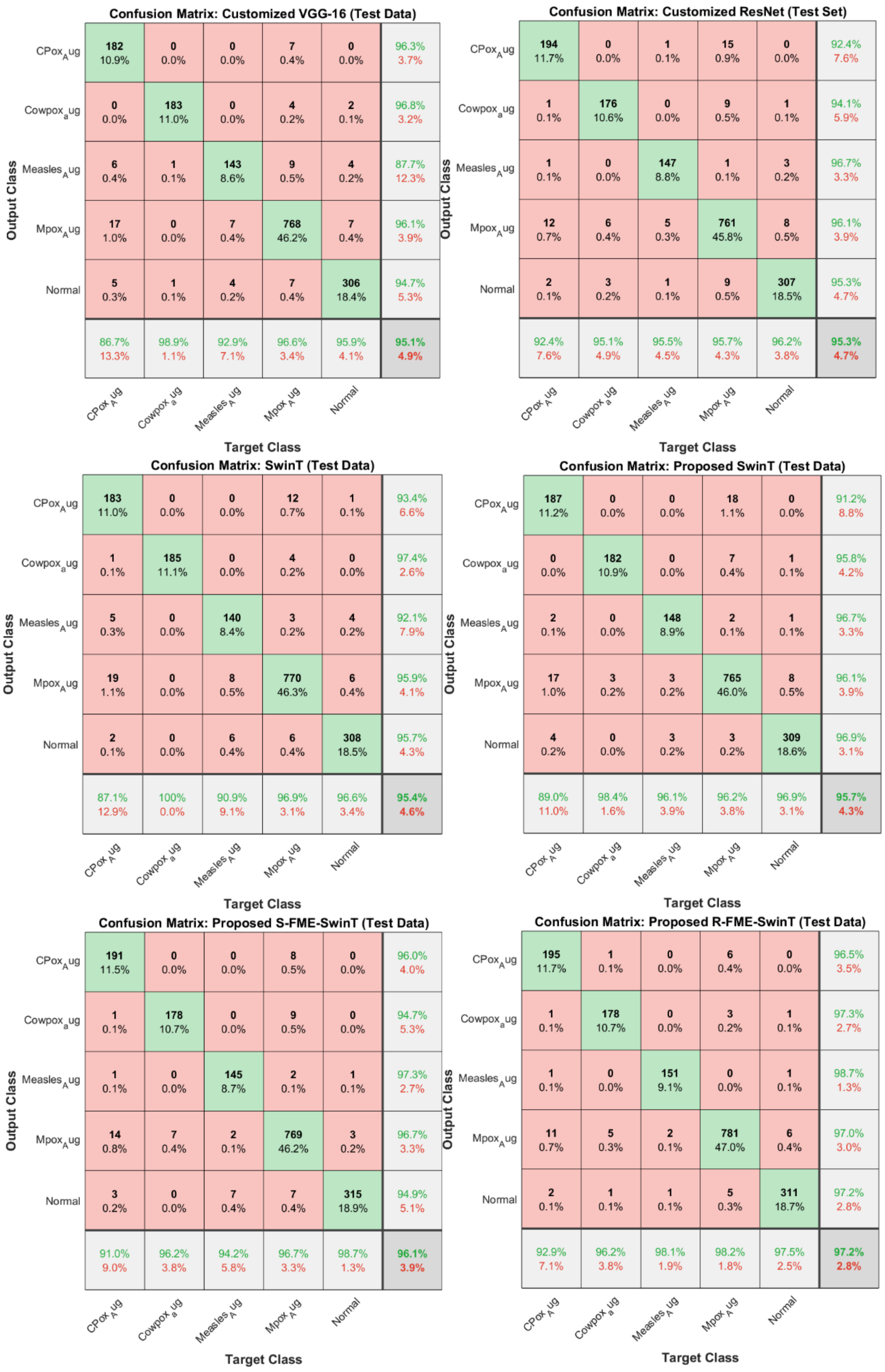

Confusion Matrix: Customized VGG-16 (Test Data)
Confusion Matrix: Customized ResNet (Test Set)
Confusion Matrix: SwinT (Test Data)
Confusion Matrix: Proposed SwinT (Test Data)
Confusion Matrix: Proposed S-FME-SwinT (Test Data)
Confusion Matrix: Proposed R-FME-SwinT (Test Data)
Output Class
Target Class
CPox_Aug
Cowpox_aug
Measles_Aug
Mpox_Aug
Normal

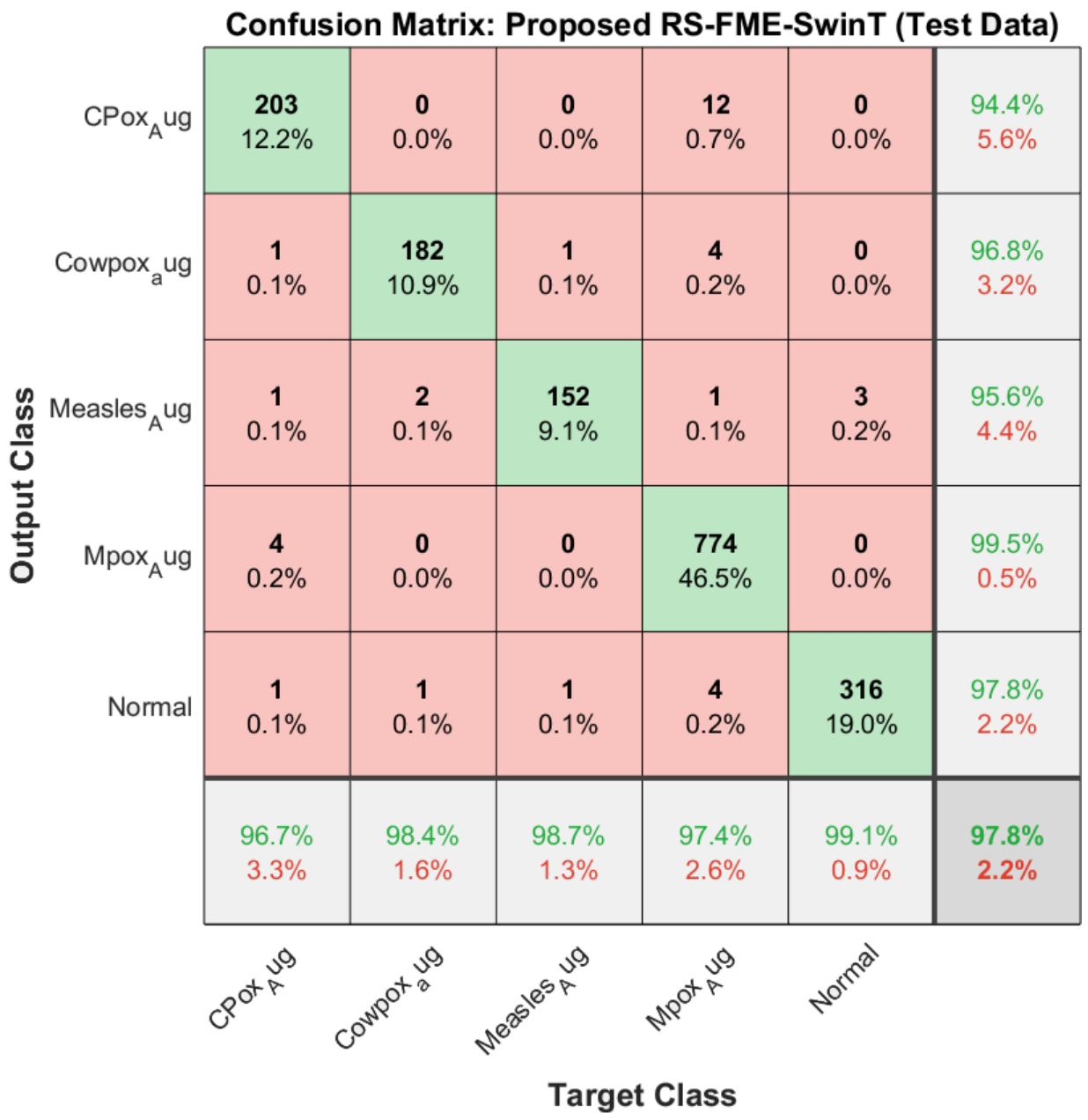

Figure 7. Confusion Matrix of the proposed techniques (RS-FME-SwinT) and existing ViTs/CNNs.

### 5.1.Performance Analysis of the Proposed RS-FME-SwinT

This study presents the RS-FME-SwinT technique for optimized MPox diagnosis. The customized SwinT extracts global features, while CNN refines texture and local features using residual learning and spatial blocks. This fusion prevents vanishing gradients and addresses illumination variations, enhancing feature capture from the MPox dataset. A feature map boosting technique merges residual learning and spatial exploitation with customized SwinT, further improving global, texture, and local feature extraction.

The SwinT employs transformer blocks to handle image patches as token sequences, capturing long-range dependencies and comprehensive context within image regions. This proves advantageous for intricate vision tasks. SwinT's scalability allows expansion by adjusting transformer layers, tokens, or hidden dimensions to capture inter-class variations. The customized SwinT achieved an accuracy of 95.71%, a sensitivity of 95.35%, precision of 95.32%, and an F-score of 95.34%, as shown in Figure 7 and Table 4. However, it may not excel at extracting local or fine-grained details for intra-class variation.

Integrating SwinT with the CNN spatial block enhances local feature extraction and global context awareness, yielding an accuracy of 96.09%, sensitivity of 95.36%, precision of 95.92%, and an F-score of 95.64%. To address vanishing gradient issues and improve texture feature extraction, integrating SwinT with a CNN residual block, which includes skip connections, improves gradient flow and prevents vanishing gradients. This integration resulted in an accuracy of 97.17%, a sensitivity of 96.57%, a precision of 97.34%, and an F-score of 96.95%, augmenting texture feature extraction and global context.

### 5.1.1. Significance of Auxiliary Channels

Finally, the proposed RS-FME-SwinT is a technique designed to leverage the integration of the strengths of CNN Spatial and Residual learning and customized SwinT for MPox disease detection. These blocks exhibit diverse abilities to learn features, utilizing various channels to capture information at different levels, revealing distinct patterns indicative of class-specific features [61]. In this RS-FME-SwinT, an ensemble learner reaches the final decision by analyzing numerous patterns extracted from the image [62]. The proposed contributes to enhanced texture and local feature extraction, along with improved global context and diverse information. The proposed RS-FME-SwinT showcases commendable results with an accuracy of 97.80%, sensitivity of 96.82%, precision of 98.06%, and an F-score of 97.44% in MPox detection, as shown in Figure 7 and Table 4.

## 5.2.Performance Gain over Existing CNNs and ViTs

This study presents a comprehensive analysis of the proposed RS-FME-SwinT and existing techniques tailored for MPox detection and evaluated on Kaggle datasets. Various CNNs and ViTs are compared, including SqueezeNet, GoogleNet, Tiny ViT, EfficientNetB0, Cross ViT, LocalViT, Xception, MobileNetV2, InceptionV3, and LeViT (LeNet), among others, applied for MPox detection Figure 8 presents our model's performance in multi-classification tasks compared to previous advanced DL methods on identical datasets.

The proposed RS-FME-SwinT demonstrated significant enhancements over traditional local correlated feature learning CNNs, with increases ranging from 3.03% to 11.90% in accuracy, 2.76% to 14.40% in sensitivity, 3.78% to 11.28% in precision, and 3.27% to 12.89% in the F1-score (Figure 9). Moreover, the technique surpassed existing global receptive learning ViTs, showing improvements in accuracy by 4.84% to 7.72%, sensitivity by 5.45% to 6.92%, precision by 5.64% to 10.61%, and the F1-score by 5.55% to 8.78%. Additionally, the proposed model demonstrates notable performance gains over hybrid CNN-ViTs (LeViT) Moreover, a comparative analysis of the most recent previous techniques employed on various datasets is mentioned in Table 5.

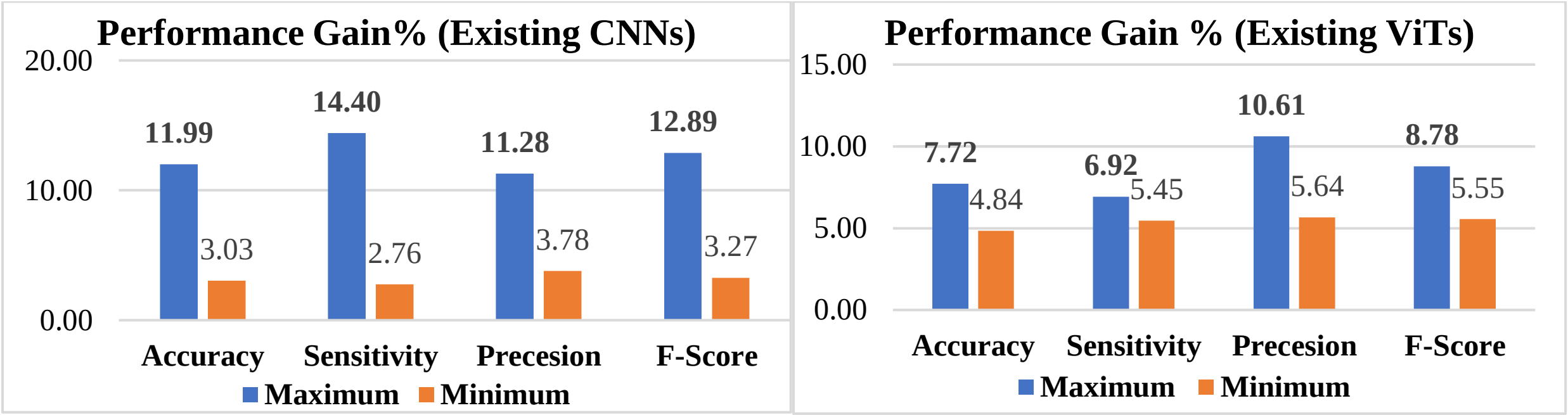

Figure 8. Performance Gain of the proposed model over existing CNNs and ViTs Models

Table 5. Performance Analysis of the existing DL models.

| Ref: | Methods | Accuracy | Precision | Sensitivity | F-Score |
|---|---|---|---|---|---|
| Sahin et al. (2022) [39] | MobileNetv2 | 91.11 | 90.00 | 90.00 | 90.00 |
| Haque et al. (2022) [42] | Xception-CBAM-Dense | 83.90 | 90.70 | 89.10 | 90.10 |
| Ali et al. (2022) [63] | ResNet50 | 82.96 | 87.00 | 83.00 | 84.00 |
| Kumar. V (2022) [64] | VGG-16 naïve Bayes | 91.11 | - | - | - |
| Sitaula et al. (2022) [30] | 13 deep learning models. Ensemble approach | 87.13 | 85.44 | 85.47 | 85.4 |
| Islam et al. (2022) [65] | ShuffleNet-V2 | 79.00 | 79.00 | 58.00 | 67.00 |
| Sizikova et al. (2022) [66] | VGG-16, EfficientNet-B3 | 88.64 | - | - | - |
| Saleh et al. (2022) [67] | Various machine learning and deep learning models | 98.48 | 91.11 | 89.00 | - |
| D Kundu et al. (2022) [45] | RestNet50 with TL | 91.00 | 91.00 | 90.00 | 90.00 |
| | ViT | 93.00 | 93.00 | 91.00 | 92.00 |
| E. H. I. Eliwa (2023) [68] | CNNs with GWO algorithm | 95.30 | 95.60 | 98.10 | 96.80 |
| T.Nayak et al. (2023) [69] | ResNet18 | 99.49 | - | - | - |
| | ViTB18 | 71.55 | 49.77 | 79.26 | 61.11 |
| A. H. Alharbi et al. (2023) [70] | GoogleNet and Metaheuristic Optimization | 94.35 | - | 95.00 | 92.00 |
| Altun, M et al. (2023) [71] | optimized hybrid MobileNetV3-s | 96.00 | - | 97.00 | 98.00 |
| A S Azar (2023) [72] | DenseNet201 | 95.18 | - | 89.82 | 89.61 |
| Aloraini (2023) [46] | ViT | 94.69 | 95.00 | 95.00 | 95.00 |
| D. Biswas et al. (2024) [73] | DarkNet53 | 85.78 | 86.92 | 82.46 | 84.20 |
| M.M. Ahsan et al. (2024) [74] | M-ResNet50 | 89.00 | 83.00 | 84.00 | 82.00 |
| Raha (2024) [75] | Attention-based MobileNetV2 | 92.28 | 90.48 | 89.42 | 89.84 |
| G Y Oztel (2024) [76] | Bagging-Ensemble ViT with Densenet201 | 81.91 | 87.17 | 74.14 | 78.16 |
| Arshed et al. (2024) [77] | ViT | 93.00 | 93.00 | 93.00 | 93.00 |

### 5.3. Ablation Study

An ablation study, employing varied block configurations, is presented in Table 4. Existing CNN spatial exploitation techniques encounter difficulties in capturing global features and extracting texture details, while residual learning faces challenges in capturing global features and minor contrast variations. Moreover, the proposed SwinT also lacks efficient feature extraction capabilities. In this regard, the proposed RS-FME-SwinT addresses these limitations, capturing multi-scale global and local correlated features for MPox diagnosis. To mitigate vanishing gradient issues in deep architectures, an improved IRB is incorporated into the SwinT and residual learning. Consequently, RS-FME-SwinT demonstrates notable performance gains over hybrid CNN- ViTs, specifically LeViT, with improvements in accuracy (2.97%), sensitivity (2.30%), precision (4.59%), and F-score (3.45%). as illustrated in Figure 7 in confusion matrix and Table 4.

Furthermore, incorporating a residual block in the customized SwinT captures local correlated features and texture features, enhancing performance with accuracy improving from 95.71% to 97.17%, sensitivity from 95.35% to 96.57%, precision from 95.32% to 97.34%, and F-score from 95.34% to 96.95%. Additionally, integrating a spatial block in the customized SwinT captures local and minor contrast variations, further improving performance with accuracy increasing from 95.71% to 96.09%, sensitivity from 95.35% to 95.36%, precision from 95.32% to 95.92%, and F-score from 95.34% to 95.64%, outperforming the existing SwinT. These analyses highlight the superiority of RS-FME-SwinT across various performance metrics.

### 5.4.Feature Space Analysis

The proposed RS-FME-SwinT technique offers an intelligible tool for enhanced comprehension and streamlined decision-making through its acquired feature representation. Figure 9 illustrates the representation of the initial two principal components (PC) within the feature space, as learned by RS-FME-SwinT, in comparison to ViT and CNN for the testing dataset. The 2D plots demonstrate that the proposed RS-FME-SwinT technique exhibits superior discriminative capability in segregating MPox-positive and non-MPox (Measles, Cowpox, Chickenpox, normal) cases compared to ViT and CNN on the test datasets. Comparative analyses against state-of-the-art ViT and CNN-based methods, including ROC-PR curves and Principal component analysis, are detailed in Figure 9. Furthermore, Figure 9 displays the generated feature instances by RS-FME-SwinT for individuals infected with MPox and provides a visual representation of the distinctive features captured by the model.

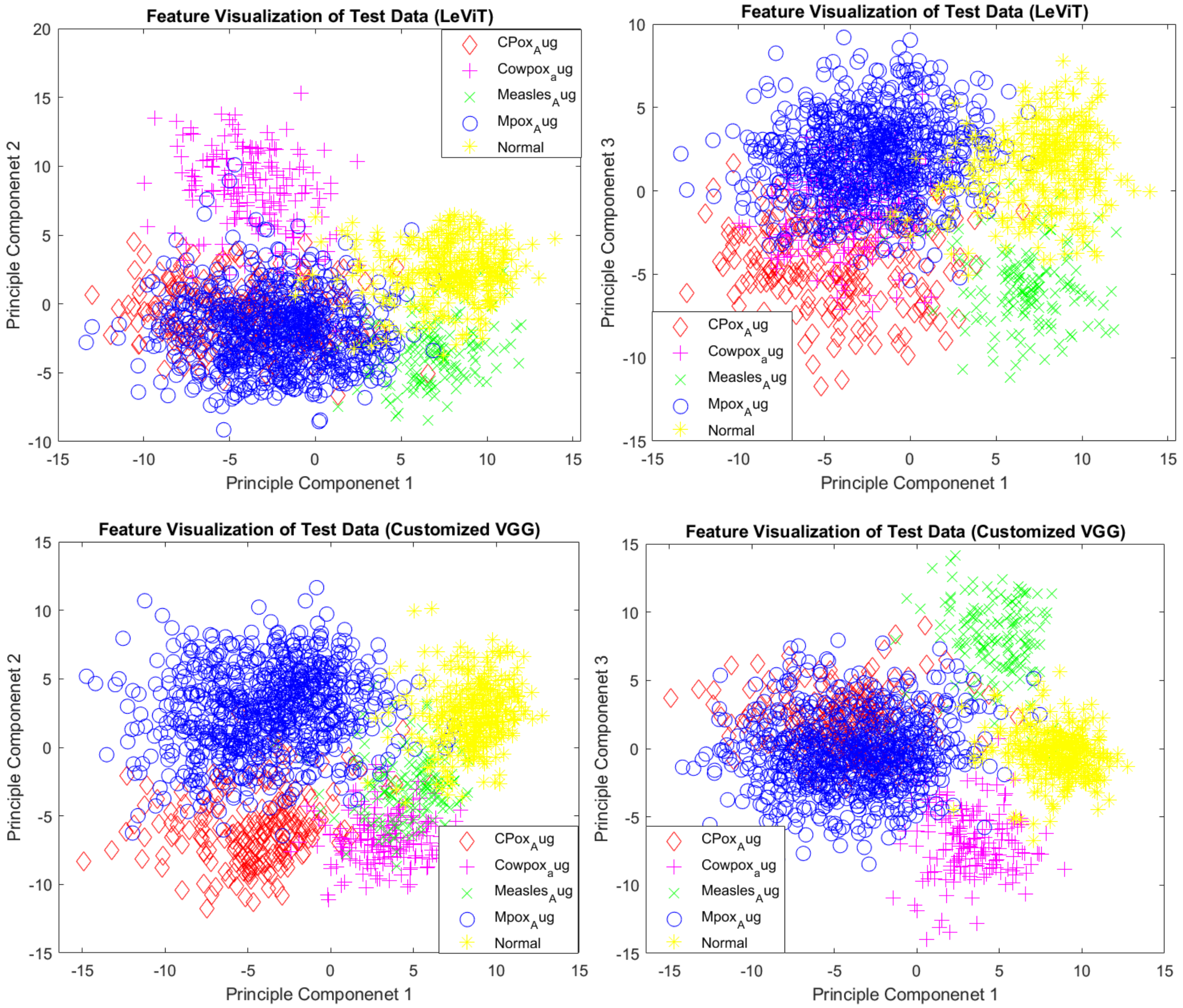

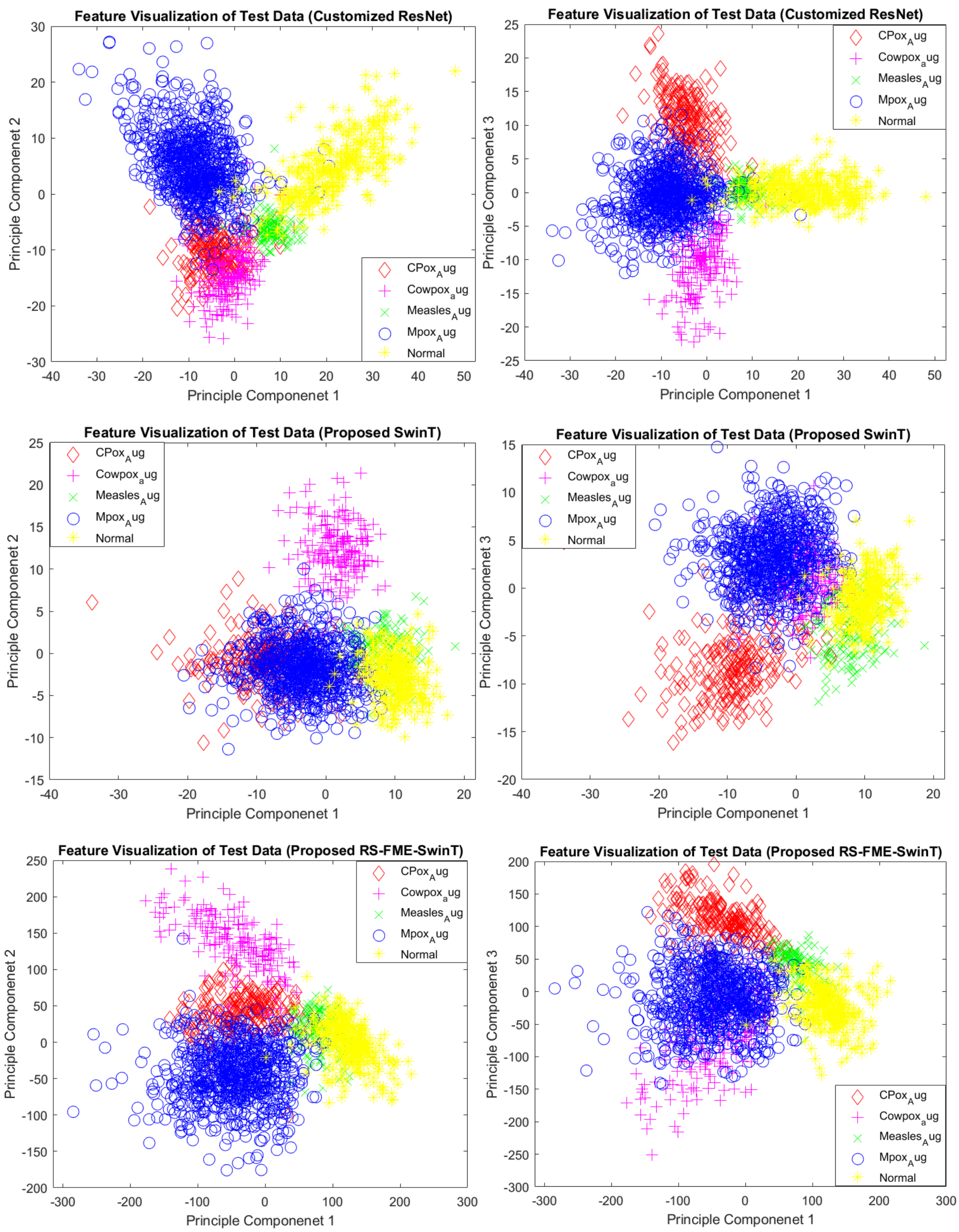

Figure 9. Feature Space Visualization of the proposed techniques (RS-FME-SwinT) on test samples.

## 5.5. Detection Rates Analysis

A robust detection rate is essential for effective MPox identification, preventing infection spread and enabling prompt treatment. RS-FME-SwinT consistently achieves the highest detection rates, ranging from 97.44% to 97.80%, with minimal FPs. The high accuracy, reaching up to 97.80%, implies a low miss-detection rate of 1% to 7%, highlighting its accurate screening ability (Table 4). This accuracy minimizes the likelihood of healthy individuals or non-MPox subjects receiving false MPox diagnoses, thus reducing the workload on healthcare professionals and scientists.

Additionally, Precision-Recall (PR) curves are crucial for determining the optimal cutoff for MPox positive and negative class (others) classification, showcasing the classifier's discrimination ability across different measurements. Figure 10 illustrates the PR curves for both the proposed and existing CNN/ViT techniques on MPox datasets. These curves demonstrate the superior performance of RS-FME-SwinT at various cutoff points. Figure 10 reveals an AUC-PR of 0.97 for the technique on MPox datasets. These high AUC values confirm that RS-FME-SwinT possesses high PR values, establishing it as a robust method for MPox detection.

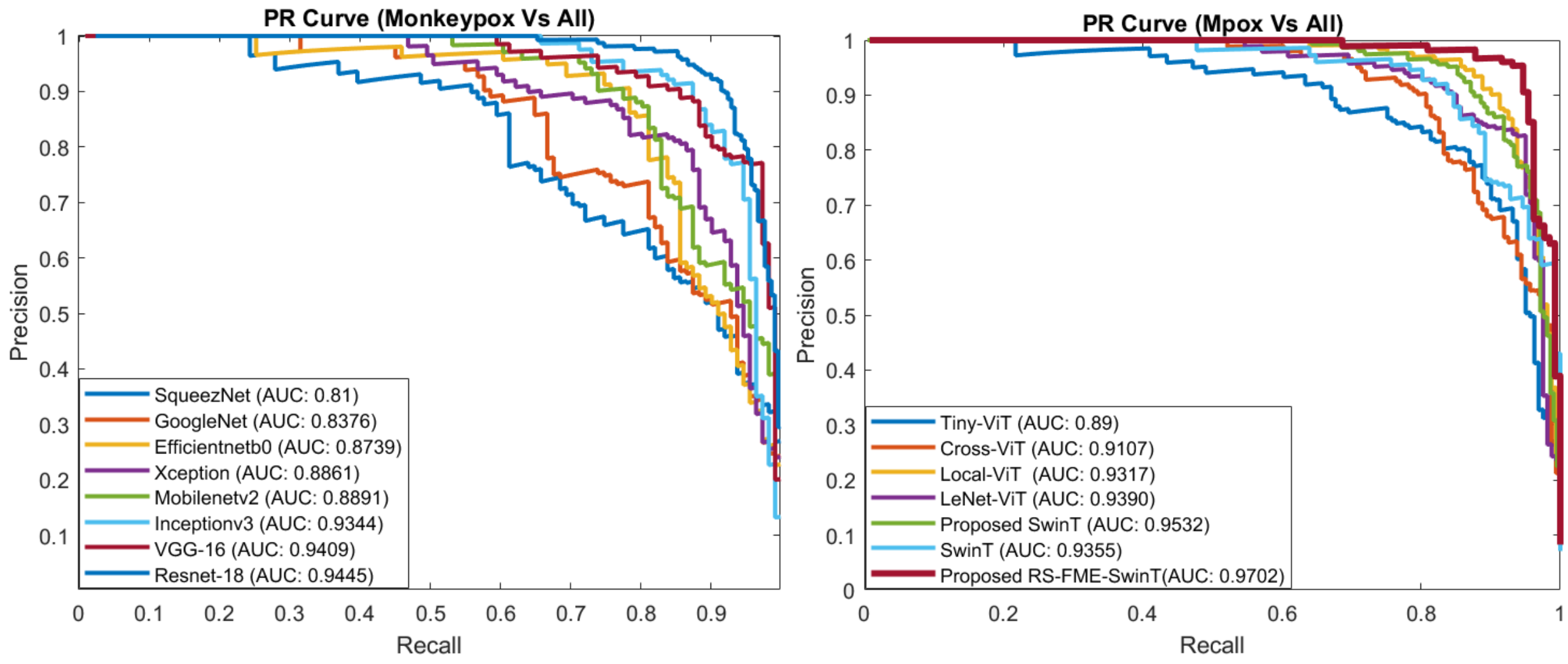

Figure 10. Detection rate analysis of the proposed (RS-FME-SwinT) and existing CNNs/ViTs.

### 5.6. Contributions to the Scientific Landscape

- The Proposed RS-FME-SwinT integrates SwinT, residual, and spatial learning, concatenating diverse channels to capture multi-level variations.
- Integrates CNNs and customized SwinT, capturing global and locally correlated features through multiple blocks. CNN channels focus on local diversity, while SwinT maps emphasize global features. Spatial heterogeneous operations address contrast and morphological variations, enhancing robustness.

- MHA and IRB in customized SwinT compute global contextual interactions and capture local texture information.
- TL-based RS-FME-SwinT preserves class-specific information at both channel and spatial levels, effectively capturing texture variation. TL and data augmentation address the challenges of limited and imbalanced data.
- Enhances the applicability of DL in diverse medical imaging tasks, improving treatment outcomes and reducing costs across various medical conditions.

## 6. Conclusion

MPox presents a significant global health challenge, requiring advanced detection methods to address its complexities. This study introduces a DL approach that effectively addresses challenges such as data scarcity, contrast variation, and lesion variability. The RS-FME-SwinT model integrates TL-based FME with a customized SwinT and CNN, enhancing feature extraction and classification accuracy. The model demonstrates strong learning capabilities, reducing intra-class MPox variation and enabling precise differentiation from other skin diseases. The proposed technique combines global information extraction through SwinT with texture and local texture pattern recognition via residual and spatial blocks, and achieves notable accuracy metrics: 97.80% accuracy, 96.82% sensitivity, 98.06% precision, and a 97.44% F-score. These outcomes confirm the model's effectiveness in real-world applications, underscoring its ability to discriminate MPox from other skin diseases with precision. This research advances AI-driven medical diagnostics and lays the groundwork for future innovations. Refining the model's approach to inter- and intra-class variability through a two-phase categorization process and enhancing region homogeneity and structural variation analysis are recommended. The proposed approach's integration into a mobile application offers significant potential to assist medical professionals in diverse settings. Additionally, exploring the ensemble of DL techniques presents a promising avenue for further enhancing diagnostic accuracy and robustness. Future research could focus on collecting more multi-modal data or using data imputation techniques, such as GANs, to improve model performance. Additionally, integrating the proposed dataset with recent datasets could enhance the framework's performance and generalization. However, handling extensive datasets may require addressing computational demands.

This manuscript includes the following abbreviations.

| Full Form | Abbreviation | Full Form | Abbreviation |
|---|---|---|---|
| Monkeypox | MPox | Multi-Head Self-Attention | MHA |
| Convolutional Neural Network | CNN | True positive | TP |
| Vision Transformer | ViT | True negative | TN |
| Swin Transformer | SwinT | False-negative | FN |
| Feature Map Enhancement | FME | False positive | FP |
| Deep Learning | DL | Data Augmentation | DA |
| Transfer Learning | TL | Feedforward Network | FFN |
| Inverse Residual Blocks | IRB | Area Under Curve | AUC |
| Monkeypox Skin Lesion Dataset | MSLD | Confidence Interval | CI |
| Monkeypox Skin Image Dataset | MSID | Support Vector Machine | SVM |
| Polymerase Chain Reaction | PCR | K-Nearest Neighbors | K-NN |

**Acknowledgment:**

We thank the Artificial Intelligence Lab, Department of Computer Systems Engineering, University of Engineering and Applied Sciences (UEAS), Swat, for providing the necessary resources.

**Conflicts of interest:**

The authors declare that they have no known competing financial interests or personal relationships that could have appeared to influence the work reported in this paper.

**Data Availability Statement**

Correspondence and requests for materials should be addressed to Saddam Hussain Khan.